\documentclass[aps]{revtex4-1}
\usepackage{graphicx}            
\usepackage{float}
\newcommand{\bea}{\begin{eqnarray}}
\newcommand{\eea}{\end{eqnarray}}
\newcommand{\avg}[1]{\ensuremath{\langle #1 \rangle}}	
\begin{document}
\title{Expansion of the conditional probability function in a network
with nearest--neighbour degree correlations}
\author{Murray E.~Alexander}
\email{Murray.Alexander@nrc-cnrc.gc.ca}
\affiliation{Institute for Biodiagnostics, National Research Council of Canada,
Winnipeg, Manitoba, R3B 1Y6, Canada \\ and \\
Physics Department, University of Winnipeg, Winnipeg, Manitoba, R3B 2E9, Canada
}
\author{Randy Kobes}
\email{r.kobes@uwinnipeg.ca}
\affiliation{Physics Department, University of Winnipeg, Winnipeg, Manitoba, R3B 2E9, Canada
\\
and\\
Winnipeg Institute for Theoretical Physics,
Winnipeg, Manitoba, R3B 2E9, Canada}
\begin{abstract}
A useful property of a network that can be used to characterize many systems is 
the degree distribution. However, many complex networks exhibit higher--order degree
correlations that must be studied through other means, such as clustering coefficients,
the Newman $r$ factor, and the average nearest neighbour degree (ANND). In this paper
we develop an expansion of the conditional probability that can be used to parameterize
such degree correlations. The measures of degree correlations associated with this
expansion can be used to signal the presence of non--linear correlations.
\end{abstract}
\maketitle
\section{Introduction}
\label{intro}
Interest in the applications of networks to technological, biological, and social
systems has grown significantly in recent years \cite{strogatz, albert, newman}.
One property of networks that has received much attention is the degree distribution,
which characterizes the number of edges that connect a given node.
Many networks show a power-law form of the degree distribution, while other networks may have an
exponential or truncated power-law distribution.
However, there is a further property of networks that is not completely captured by the
degree distribution alone: it is found that
most networks have a high degree of clustering, with nodes
tending to share many common connections.
\par
A number of ways have been developed to characterize such correlations.
One standard approach was formulated by
Newman \cite{newman1, newman2, newman3, weber}, who introduced what is now
called the Newman factor $r$. 
This number is essentially the Pearson correlation coefficient of degrees 
from connected vertices in a network
 and is fully defined by two--point correlations in a network. 
 The Newman factor is normalized to lie in the interval $-1 \le r \le 1$,
 and is defined so that positive (negative) values indicate that vertices with 
 the same (different) degree tend to be connected, which indicate assortative (disassortative)
 mixing. A Newman factor of 0 means no correlations are present. 
 Most networks have a non-trivial Newman factor: 
 biological networks tend to show negative $r$ values, 
technological networks display values close to zero, 
while social networks tend to have rather large positive values \cite{newman}.
Another, related measure of correlations that is commonly used is the
average nearest neighbour degree  (ANND) ${\bar k}_{nn}(k)$, which is a function in
principle of the degree $k$ \cite{annd}. For uncorrelated networks, the ANND is 
independent of $k$, and thus an explicit $k$ dependence indicates the presence of
degree correlations. A ${\bar k}_{nn}(k)$ that is an increasing function of
$k$ shows assortative mixing, while a decreasing function of $k$ indicates
disassortative mixing. A third commonly used measure of correlations are
clustering coefficients \cite{albert}, which can be related to these other
measures under certain assumptions \cite{boguna, dorogovtsev}.
\par
However, some of these measures may not completely capture all correlations that
may be present in a network. For example, the Pearson correlation coefficient 
by definition only explores linear relationships
between two variables \cite{stats}.  In this paper, we develop a formalism that in principle
can be used to measure higher--order non--linear correlations. We do this by examining
an expansion of $P(k, k^\prime)$, the joint probability that an edge chosen at random
connects vertices
of degree $k$ and $k^\prime$. In Section \ref{expansion} we develop this expansion,
through which we can define generalizations of the Newman $r$ factor and the
ANND ${\bar k}_{nn}(k)$ which are sensitive to non--linear correlations. 
The expansion developed in this section involves certain input coefficients;
in Section \ref{sannd} we show how to relate this expansion to one involving
the specification of certain generalized ANND functions.
In Section \ref{clustering}
we use the results of Ref.\cite{dorogovtsev} to show how these arise in
defining clustering coefficients. In Section \ref{examples} we illustrate the use and
effects of these generalizations in some simple examples. Section \ref{conclusions} contains
some brief conclusions.
\section{Expansion of the conditional probability function}
\label{expansion}
\par
Let $P(k)$ denote the probability of finding a node of degree $k$,
$P(k^\prime | k)$ the conditional probability that a vertex of degree $k$ 
is connected to a vertex of degree $k^\prime$,
and $P(k, k^\prime) = P(k^\prime, k)$ 
be the joint probability that a randomly chosen edge connects vertices
of degree $k$ and $k^\prime$. These satisfy the normalization conditions
\begin{equation}
\sum_k P(k) = \sum_{k^\prime} P(k^\prime | k) = 1
\label{normalize}
\end{equation}
as well as the detailed balance equation
\begin{equation}
k P(k^\prime | k) P(k) = k^\prime P(k|k^\prime) P(k^\prime)
\label{detbal1}
\end{equation}
Let us now introduce the edge distribution \cite{weber}
\begin{equation}
P_e(k) = \sum_{k^\prime} P(k^\prime, k) 
\end{equation}
This allows us to express the joint probability $P(k^\prime, k)$ to the conditional
probability $P(k^\prime | k)$ as
\begin{equation}
P(k^\prime, k) = P(k^\prime | k) P_e(k)
\label{jtoc} \end{equation}
If we now define, for any function $f(k)$,
\begin{eqnarray}
 {\overline { f(k)} } &=& \sum_k f(k) P(k)\nonumber\\
 \avg{f(k)} &=& \sum_k f(k) P_e(k)
 \end{eqnarray}
 we can relate the edge distribution $P_e(k)$ to the degree distribution $P(k)$ as
 \begin{equation}
 P_e(k) = \frac{k}{ {\overline k} } P(k)
 \end{equation}
 In terms of $P_e(k)$, the detailed balance condition of Eq.~(\ref{detbal1}) becomes
\begin{equation}
P(k^\prime | k) P_e(k) = P(k|k^\prime) P_e(k^\prime)
\label{detbal}
\end{equation}
\par
We now examine the form of the conditional probability $P(k^\prime | k)$.
First note that, if the network were uncorrelated, we would have
\begin{equation}
P_{nc}(k^\prime | k) = \frac{ k^\prime P(k^\prime) } { \overline k } = P_e(k^\prime)
\end{equation}
This suggests that we can put $P(k^\prime | k)$ into the form
\begin{equation}
P(k^\prime | k) = P_e(k^\prime) + Q(k^\prime | k)
\label{ansatz1}
\end{equation}
for some function $Q(k^\prime | k)$.
The detailed balance condition of Eq.~(\ref{detbal}) implies
\begin{equation}
Q(k^\prime | k) P_e(k) = Q(k | k^\prime) P_e(k^\prime)
\Rightarrow Q(k^\prime | k) = P_e(k^\prime) Q(k^\prime, k)
\end{equation}
where $Q(k^\prime, k) = Q(k, k^\prime)$ is symmetric in $k$ and $k^\prime$.
With this, we can then put $P(k^\prime | k)$ of Eq.~(\ref{ansatz1}) into the form
\begin{equation}
P(k^\prime | k) = P_e(k^\prime)\left[ 1 + Q(k^\prime , k)\right]
\label{ansatz}
\end{equation}
subject to the condition
\begin{equation}
\sum_{k^\prime} Q(k^\prime , k)P_e(k^\prime) = 0 
\label{constraint}
\end{equation}
which follows from Eq.~(\ref{normalize}).
\par
Let us assume that we can expand the function
$Q(r, s)$ in in terms of symmetric polynomials of $r$ and $s$:
\begin{equation}
Q(r, s) = E_0(r, s) + E_1(r, s)+E_2(r, s)+E_3(r, s)+E_4(r, s)+\ldots
\end{equation}
where
\begin{eqnarray}
E_0(r, s) &=& e_{00}^{(0)} \nonumber\\
E_1(r, s) &=& e_{11}^{(1)} (r+s)  \nonumber\\
E_2(r, s) &=& e_{22}^{(2)} (r^2+s^2) + e_{11}^{(2)} rs \nonumber\\
E_3(r, s) &=& e_{33}^{(3)} (r^3+s^3) + e_{21}^{(3)} (r^2s+rs^2) \nonumber\\
E_4(r, s) &=& e_{44}^{(4)} (r^4+s^4) + e_{31}^{(4)} (r^3s+rs^3)
+ e_{22}^{(4)} r^2s^2 
\end{eqnarray}
and $e_{ij}^{(k)}$ are constant coefficients.
The constraint of Eq.~(\ref{constraint}) will lead to conditions on certain coefficients of
these expansions to a given order. This will allow us to
 write the expansion of $Q(m, n)$ to the $t^{\rm th}$ order as
\begin{equation}
Q^{(t)}(r, s) = \sum_{i=1}^t \sum_{j=1}^t \alpha_{ij}\, v_i(r)  v_j(s)
\label{qv}
\end{equation}
where we have introduced
\begin{eqnarray}
v_n(k) &=& \frac{1}{\sqrt{\sigma_{nn}}} \left[ k^n-\avg{k^n}\right] \nonumber\\
\sigma_{ij} &\equiv& \avg{ (k^i - \avg{k^i}) (k^j - \avg{k^j}} = 
\avg{ k^i  (k^j - \avg{k^j})} = \avg{k^{i+j}} - \avg{k^i}\avg{k^j}
\label{v}\end{eqnarray}
Note that $\avg{v_i} = 0$ and 
$\avg{v_i v_i} = 1$ for $i=1,2, \ldots, n$, but $\avg{v_i v_j} \neq 0$ for $i \neq j$.
Through the Gram--Schmidt procedure, we can develop an orthonormal basis $u_i(k)$ 
satisfying $\avg{u_i} = 0$  and $\avg{u_i u_j} = \delta_{ij}$ from
the $v_i(k)$ as
\begin{equation}
u_i(k) = \frac{ U_i(k)}{ \sqrt{\avg{U_i U_i} }}
\label{unorm}
\end{equation}
where
\begin{eqnarray}
U_1(k) &=& v_1(k) \nonumber\\
U_2(k) &=& v_2(k) -  \avg{u_1 v_2} u_1(k) \nonumber\\
U_3(k) &=& v_3(k) - \avg{u_1 v_3}  u_1(k) 
 -  \avg{u_2 v_3}  u_2(k)  \nonumber\\
&\vdots& \nonumber\\
U_n(k) &=& v_n(k) - \sum_{j=1}^{n-1} \avg{u_j v_n} u_j(k)
\label{gm}
\end{eqnarray}
The relations of Eqs.(\ref{unorm}, \ref{gm}) allow us to express the $v_i(k)$ in
terms of the $u_i(k)$:
\begin{equation}
v_n(k) = \sum_{j=1}^n \avg{u_j v_n} u_j(k)
\label{gma}\end{equation}
This allows us to express the expansion of $Q(s, t)$ of
Eq.~(\ref{qv}) as
\begin{eqnarray}
Q^{(t)}(r, s) &=& \sum_{a=1}^t \sum_{b=1}^t \alpha_{ab}\, v_a(r) v_b(s)\nonumber\\
&=&  \sum_{a=1}^t \sum_{b=1}^t \sum_{i=1}^a \sum_{j=1}^b \alpha_{ab}\, 
\avg{u_i v_a} \avg{u_j v_b} u_i(r) u_j(s) \nonumber\\
&=&  \sum_{a=1}^t \sum_{b=1}^t \sum_{i=1}^t \sum_{j=1}^t  \alpha_{ab}\, 
\avg{u_i v_a} \avg{u_j v_b} \theta(a-i) \theta(b-j)  u_i(r) u_j(s) \nonumber\\
&=& \sum_{i=1}^t \sum_{j=1}^t \beta_{ij} \, u_i(r) u_j(s) 
\label{qu}
\end{eqnarray}
where $\theta(x) = 1$ if $x \ge 1$ and zero otherwise, and
\begin{equation}
\beta_{ij} =  \sum_{a=1}^t \sum_{b=1}^t  \alpha_{ab}\, 
\avg{u_i v_a} \avg{u_j v_b} \theta(a-i) \theta(b-j)
\end{equation}
\par
The coefficients $\beta_{ij}$ can be interpreted in terms of generalized
correlation coefficients as follows.
We have, from Eq.~(\ref{qu}) and the fact that $\avg{u_i u_j} = \delta_{ij}$, the relationship
\begin{equation}
\sum_{r} \sum_{s} u_a(r) u_b(s) Q^{(t)}(r, s) P_e(r) P_e(s) = \beta_{ab}
\label{r1}
\end{equation}
We now introduce a generalized $a^{\rm th}$--order ANND $k_{nn}^{(a)}(k)$ as
\begin{equation}
 k_{nn}^{(a)}(k)  = \sum_{k^\prime} u_a(k^\prime) P(k^\prime | k)
 = \sum_{k^\prime} u_a(k^\prime) Q^{(t)}(k^\prime, k) P_e(k^\prime)
 = \sum_{j=1}^t \beta_{aj} u_j(k)
 \label{annd}
 \end{equation}
 where we have used Eqs.(\ref{ansatz}, \ref{qu}) and $\avg{u_i(k)} = 0$. This allows us to write
 $Q^{(t)}(r, s)$ of Eq.~(\ref{qu}) as
 \begin{equation}
 Q^{(t)}(r, s) = \sum_{i=1}^t u_i(r) k_{nn}^{(i)}(s) = \sum_{i=1}^t k_{nn}^{(i)}(r) u_i(s) 
\label{qt}  \end{equation}
 and also allows us to express 
 Eq.~(\ref{r1}) as
\begin{equation}
\beta_{ab} = \avg{ k_{nn}^{(a)}(k) u_b(k) }
\end{equation}
We can then use this to introduce
generalized correlation coefficients $r_{ab}$ to order $t$:
\begin{equation}
r_{ab} = \frac{
          \avg{   
                   ( k_{nn}^{(a)}  - \avg{ k_{nn}^{(a)} } )
                   (u_b -\avg{u_b} )  
           } 
      } 
      {  
         \sqrt{   
              \avg{ 
                    (k_{nn}^{(a)} - \avg{k_{nn}^{(a)} } )^2
                } 
           } 
           \sqrt{
                \avg{
                    (u_b - \avg{u_b} )^2
                 } 
            } 
         }
  = \frac{ \avg{  k_{nn}^{(a)} u_b } }
  { \sqrt{ \avg{ (k_{nn}^{(a)} )^2 } } } 
 \equiv
 \frac{\beta_{ab}}{R_a}
 \label{correlation}
 \end{equation}
where $a, b=1, 2, \ldots, t$, 
 we have used $\avg{k_{nn}^{(a)}} = 0 = \avg{u_i}$, and we have defined
 \begin{equation}
R_a =  \sqrt{ \avg{ (k_{nn}^{(a)} )^2  }  }= \sqrt{ \sum_{j, k=1}^t \beta_{aj} \beta_{ak} \avg{ u_j u_k} } =
 \sqrt{ \sum_{j, k=1}^t \beta_{aj} \beta_{ak} \delta_{jk} } = \sqrt{ \sum_{j=1}^t \beta_{aj}^2 }
 \end{equation}
 The definition of $r_{ab}$ in Eq.~(\ref{correlation})
satisfies $-1 \le r_{ab} \le 1$, which
follows from the  Cauchy--Schwarz inequality
\begin{equation}
\left| \avg{ (X-\avg{X})(Y-\avg{Y}) } \right|^2 \le \avg{ (X-\avg{X})^2} \avg{ (Y-\avg{Y})^2} 
\end{equation}
for two variables $X$ and $Y$. We can then write the expansion of $Q^{(t)}(r, s)$ of Eq.~(\ref{qu})
as
\begin{equation}
Q^{(t)}(r, s) = \sum_{a=1}^t \sum_{b=1}^t R_a r_{ab} u_a(r) u_b(s)
\label{polar}\end{equation}
with the condition $R_a r_{ab} = R_b r_{ba}$ coming from $\beta_{ab} = \beta_{ba}$.
Note that, to a given order $t$, the $r_{ab}$ are not independent, since by Eq.~(\ref{correlation})
we have $r_{a1}^2 + r_{a2}^2 + \ldots + r_{at}^2 = 1$. This allows us to 
introduce $t-1$ hyperspherical angles $\theta_{ak}$ and parameterize the
$r_{ab}$ as
\begin{equation}
\displaystyle
\left( \begin{array}{c}
r_{a1}\\ r_{a2} \\ r_{a3} \\ \vdots \\ r_{ak} \\ \vdots \\ r_{a(t-1)} \\ r_{at}
\end{array} \right) = 
\frac{1}{R_a}
\left( \begin{array}{c}
\beta_{a1}\\ \beta_{a2} \\ \beta_{a3} \\ \vdots \\ \beta_{ak} \\ \vdots \\ \beta_{a(t-1)} \\ \beta_{at}
\end{array} \right) = 
 \left( \begin{array}{c}
\cos\theta_{a1} \\ \sin\theta_{a1}\cos\theta_{a2} \\ \sin\theta_{a1}\sin\theta_{a2}
  \cos\theta_{a3} \\ \vdots \\
\left( \Pi_{i=1}^{k-1} \sin\theta_{ai} \right) \cos\theta_{ak} \\ \vdots \\
\sin\theta_{a1}\cdots\sin\theta_{a(t-2)}\cos\theta_{a(t-1)} \\
\sin\theta_{a1}\cdots\sin\theta_{a(t-2)}\sin\theta_{a(t-1)}
\end{array} \right)
\end{equation}
This form of the $r_{ab}$, as appearing in the expansion of Eq.~(\ref{polar}),
makes it explicit that it is the combination of the $\beta_{ab}$ parameters that appear
in $R_a$ that sets the scale for  $Q^{(t)}(r, s)$.
\par
It is straightforward to see how the usual ANND $k_{nn}(k)$, and related Newman
correlation coefficient $r$, are related to the generalized $k_{nn}^{(a)}(k)$ of Eq.~(\ref{annd})
and $r_{ab}$ of Eq.~(\ref{correlation}). Suppose we keep only the leading--order
$t=1$  term in the expansion of $Q^{(t)} (r, s)$ in Eq.~(\ref{qv}):
\begin{equation}
Q^{(1)}(r, s) = \alpha_{11} v_1(r)v_1(s)
\end{equation}
where, from Eq.~(\ref{v}),
\begin{equation}
v_1(k) = \frac{ k-\avg{k} } { \sqrt{\sigma_{11} } } =
\frac{ k - \avg{k} } { \sqrt{ \avg{k^2} - \avg{k}^2 } }
\label{v1} \end{equation}
In this case only the orthonormal basis vector $u_1(k) = v_1(k)$ of Eq.~(\ref{unorm}) would arise,
leading to the expansion
\begin{equation}
Q^{(1)}(r, s) = \beta_{11}u_1(r)u_1(s)
\end{equation}
The only generalized ANND function of Eq.~(\ref{annd}) that would appear would be
\begin{equation}
 k_{nn}^{(1)}(k)  = \sum_{k^\prime} u_1(k^\prime) P(k^\prime | k)
 = \frac{1}{\sqrt{ \sigma_{11}}} \sum_{k^\prime} (k^\prime - \avg{k^\prime} ) P(k^\prime | k)
 = \frac{1}{\sqrt{ \sigma_{11}} } \left[  \sum_{k^\prime} k^\prime P(k^\prime | k) 
 - \avg{k} \right]
 \end{equation}
 Since the standard ANND ${\bar k}_{nn}(k)$ is defined as \cite{boguna}
 \begin{equation}
 {\bar k}_{nn}(k) =  \sum_{k^\prime} k^\prime P(k^\prime | k)
 \end{equation}
which satisfies $\avg{ {\bar k}_{nn}(k) } = \avg{k}$, we then have the relationship
 \begin{equation}
  k_{nn}^{(1)}(k)  = \frac{1}{ \sqrt{ \sigma_{11}}} \left[  {\bar k}_{nn}(k)  - \avg{k} \right]
  = \frac {  {\bar k}_{nn}(k)  - \avg{ {\bar k}_{nn}(k)} } { \sqrt{ \avg{k^2} - \avg{k}^2 } }
 \label{anndrel} \end{equation}
We can also relate the standard Newman correlation factor $r$ to the corresponding
generalization of Eq.~(\ref{correlation}). The Newman factor $r$ can be written as \cite{weber}
\begin{eqnarray}
r &=& \frac{1} { \avg{k^2} - \avg{k}^2 }
\sum_{k^\prime, k} k^\prime k \left[ P(k^\prime, k) - P_e(k^\prime) P_e(k) \right] =
\nonumber\\
&=& \frac{1}{\sigma_{11} }
\left[ \sum_{k^\prime, k} k^\prime k P(k^\prime | k) P_e(k) - \avg{k}^2 \right] = 
 \frac{1}{\sigma_{11} }
\left[ \sum_k {\bar k}_{nn}(k) k P_e(k)  - \avg{k}^2 \right] = \nonumber\\
&=& \frac{1}{\sigma_{11} } 
\avg{  ( {\bar k}_{nn}(k) - \avg{ {\bar k}_{nn}(k) } ) ( k - \avg{k} ) }
\end{eqnarray}
where we have used Eq.~(\ref{jtoc}) to relate the joint probability to the conditional
probability, as well as $\avg{{\bar k}_{nn}(k)} = \avg{k}$. Using Eq.~(\ref{anndrel}) to
relate ${\bar k}_{nn}(k)$ to $ k_{nn}^{(1)}(k)$, as well as Eq.~(\ref{v1}) to relate
$k$ to $u_1(k)$, we then have
\begin{equation}
r = \frac{1}{\sigma_{11} } \left( \sqrt{\sigma_{11} } \right)^2 
\avg{ k_{nn}^{(1)}(k) u_1(k) } = \avg{ k_{nn}^{(1)}(k) u_1(k) }
\end{equation}
The corresponding correlation coefficient of Eq.~(\ref{correlation}) that would appear 
to this order is
\begin{equation}
r_{11} = \frac{
          \avg{   
                   ( k_{nn}^{(1)}  - \avg{ k_{nn}^{(1)} } )
                   (u_1 -\avg{u_1} )  
           } 
      } 
      {  
         \sqrt{   
              \avg{ 
                    (k_{nn}^{(1)} - \avg{k_{nn}^{(1)} } )^2
                } 
           } 
           \sqrt{
                \avg{
                    (u_1 - \avg{u_1} )^2
                 } 
            } 
         }
         = \frac{ \avg{ k_{nn}^{(1)} u_1 } } 
         {   \sqrt{  \avg{  (k_{nn}^{(1)})^2 } } }=  \frac{ \beta_{11} } { |\beta_{11} |}
\end{equation}
which allows us to identify $r = \beta_{11} = r_{11}  |\beta_{11} | $. Note that, to this order,
we have $r_{11} = \pm 1$, which simply reflects the fact that in the linear approximation
there is perfect linear correlation or anti--correlation between $ k_{nn}^{(1)}$ and $u_1$.
\section{Specifying the Average Nearest Neighbour Degree}
\label{sannd}
An alternate, but related, approach to the expansion discussed here has been formulated by
Weber and Porto \cite{weber}, who examined how one could expand the joint probability
in terms of a particular functional form for the ANND ${\bar k}_{nn}(k)$. In the current notation,
this approach proceeds as follows. One begins with the functional form of Eq.~(\ref{ansatz}) for the
conditional probability:
\begin{equation}
P(k^\prime | k) = P_e(k^\prime)\left[ 1 + Q(k^\prime , k)\right]
\end{equation}
subject to the constraint of Eq.~(\ref{constraint}):
\begin{equation}
\sum_{k^\prime} Q(k^\prime , k)P_e(k^\prime) = 0 
\end{equation}
One then assumes the symmetric function $Q(k^\prime, k)$ has the form
\begin{equation}
Q(k^\prime, k) = h(k^\prime) h(k)
\end{equation}
and expresses the ANND ${\bar k}_{nn}(k)$ in terms of $h(k)$ as
\begin{equation}
{\bar k}_{nn}(k) = \sum_{k^\prime} k^\prime P(k^\prime | k) =
\avg{k} + \sum_{k^\prime} k^\prime h(k^\prime) P_e(k^\prime) h(k)
= \avg{k} + \avg{k h(k)} h(k)
\end{equation}
which allows one to infer
\begin{equation}
h(k) = \frac{ {\bar k}_{nn}(k) - \avg{k} } { \avg{k h(k)} }
\end{equation}
From this, one has
\begin{equation}
\avg{ kh(k) } =  \frac{ \avg{ k {\bar k}_{nn}(k) - k \avg{k} }}  { \avg{k h(k)} }
\Rightarrow \avg{ kh(k) } = \sqrt{ \avg{k {\bar k}_{nn}(k)} - \avg{k}^2 }
\end{equation}
The function $Q(k^\prime, k)$ arising in the joint probability
\begin{equation}
Q(k^\prime, k) = 
\frac{ ({\bar k}_{nn}(k^\prime) - \avg{k} )({\bar k}_{nn}(k) - \avg{k} ) }
{ \avg{ k {\bar k}_{nn}(k) } - \avg{k}^2 }
\label{qweber} \end{equation}
is thus specified by ${\bar k}_{nn}(k)$. In order to respect the identity 
$\avg{ {\bar k}_{nn}(k) } = \avg{k}$, we can parameterize ${\bar k}_{nn}(k)$ as
\begin{equation}
{\bar k}_{nn}(k) = \frac{ \avg{k} } { \avg{ g(k) } } g(k)
\label{gweber} \end{equation}
for a suitable function $g(k)$.
\par
This approach can be related to the expansion considered here. To do this,
we first cast the expansion of $Q(k^\prime, k)$ in Eq.~(\ref{qweber}) as
\begin{equation}
Q(k^\prime, k) = \frac{ k_{nn}^{(1)}(k^\prime) k_{nn}^{(1)}(k) }
{ \avg{ k_{nn}^{(1)}(k) u_1(k) } }
\label{qweberequiv} \end{equation}
and the parameterization of ${\bar k}_{nn}(k) $ of Eq.~(\ref{gweber}) as
\begin{equation}
 k_{nn}^{(1)}(k) = \frac{ \avg{k} } { \sqrt{\sigma_{11} }} \left[
 \frac{g(k)} {\avg{g(k)} } - 1 \right]
 \end{equation}
where Eqs.(\ref{v1}, \ref{anndrel}) have been used. This is to be compared to the
general expansion of Eq.~(\ref{qt}):
\begin{equation}
 Q^{(t)}(k^\prime, k) = \sum_{i=1}^t \sum_{j=1}^t \beta_{ij} \, u_i(k^\prime) u_j(k) =
 \sum_{i=1}^t u_i(k^\prime) k_{nn}^{(i)}(k) = 
 \sum_{i=1}^t k_{nn}^{(i)}(k^\prime) u_i(k) 
 \end{equation}
 Rather than considering the $\beta_{ij}$ coefficients as input parameters to $ Q^{(t)}(k^\prime, k) $, 
 we could, in analogy with Eq.~(\ref{gweber}),  specify the ANND $k_{nn}^{(a)}(k)$ by 
 some function $f^{(a)}(k)$. 
In order to
respect the normalization $\avg{k_{nn}^{(a)}(k)} = 0$, we write
\begin{equation}
k_{nn}^{(a)}(k) = \left[ f^{(a)}(k) - \avg{ f^{(a)}(k) }  \right]
\end{equation}
Introducing 
\begin{equation}
\sigma_{f^a} = \sqrt{ \avg{ (f^{(a)}(k) - \avg{f^{(a)}(k)} )^2 } }
\end{equation}
we can write this as
\begin{equation}
k_{nn}^{(a)}(k) = \sigma_{f^a} \frac{ f^{(a)}(k) - \avg{ f^{(a)}(k) } }
{\sigma_{f^a} } \equiv \sigma_{f^a} g^{(a)}(k)
 \end{equation}
so that $g^{(a)}(k)$ has zero mean and unit norm. We then expand $g^{(a)}(k)$ in terms
of the orthonormal basis $u_a(k)$ to some order $t$:
\begin{equation}
g^{(a)}(k) = \sum_{b=1}^t c^{(a)}_b u_b(k)
\label{gex}
\end{equation}
with the coefficients $c^{(a)}_b $ given by $c^{(a)}_b  = \avg{u_b(k) g^{(a)}(k)}$. 
Comparing this to the expansion
of Eq.~(\ref{annd}) of $k_{nn}^{(a)}(k)$ in terms of the $\beta_{ij}$ allows us to identify
\begin{equation}
\beta_{ab} = \sigma_{f^a} c_b^{(a)} = \sigma_{f^a}  \avg{u_b(k) g^{(a)}(k)}
\end{equation}
In this approach, the order $t$ to which one is working should be chosen so that the expansion
in Eq.~(\ref{gex}) of $g^{(a)}(k)$ in terms of the orthonormal basis $u_i(k)$  is accurate.
\par
Note that, in analogy with the expansion of $Q(k^\prime, k)$ of Eq.~(\ref{qweberequiv}) developed by
Weber and Porto \cite{weber}, we could consider in the present approach a generalization of
the expansion of Eq.~(\ref{qt}):
\begin{equation}
Q^{(t)}(k^\prime, k) =  \sum_{i=1}^t k_{nn}^{(i)}(k^\prime) u_i(k)  \to
 \sum_{i=1}^t \frac{ k_{nn}^{(i)}(k^\prime) k_{nn}^{(i)}(k) }
 { \avg{ k_{nn}^{(i)}(k) u_i(k) } }
 \end{equation}
As with any perturbative expansion, the potential advantages of this form over the expression
of Eq.~(\ref{qt}) will depend on these higher--order terms being small, in some sense,
to the unperturbed case.
\section{Clustering Coefficients}
\label{clustering}
Clustering measures correlations among 3 nodes in a network, and so requires
knowledge of the conditional probability $P( k^\prime, k^{\prime \prime} | k)$, which is the
probability that a vertex of degree $k$ is simultaneously connected to two vertices of degree
$k^\prime$ and $  k^{\prime \prime}$ \cite{boguna}. For non--Markovian networks  
$P( k^\prime, k^{\prime \prime} | k)$ and $P(k^\prime | k)$ are unrelated, but for
Markovian networks, we have the relation 
$P( k^\prime, k^{\prime \prime} | k) = P(k^\prime | k) P(k^{\prime\prime} | k)$, so knowledge
of the two--point probability distribution is sufficient.
As shown by Dorogovtsev \cite{dorogovtsev}, various measures of clustering in a 
network can be derived in this case taking into account degree correlations present
in $P(k^\prime | k)$.
Three related clustering coefficients can be introduced:
\begin{itemize}
\item
The degree--dependent local clustering coefficient $C(k)$:
\begin{equation}
C(k) = \frac{ \avg{ m_{nn}(k) } } {k(k-1)/2}
\label{ck}\end{equation}
where $\avg{ m_{nn}(k) }$ is the average number of connections between the nearest neighbours
of a vertex of degree $k$.
\item
The mean clustering coefficient ${\overline C}$:
\begin{equation}
{\overline C} = \sum_k P(k)C(k)
\label{cm}\end{equation}
\item
The clustering coefficient $C$:
\begin{equation}
C = \frac{ \sum_k P(k) \avg{ m_{nn}(k) } } { \sum_k P(k) k(k-1)/2 } =
 \frac{ 1} { {\overline k} ( \avg{k} - 1) } \sum_k k(k-1) P(k)C(k) 
 \label{c}\end{equation}
 \end{itemize}
As shown in Ref.~\cite{dorogovtsev}, assuming the size $N$ of the network is 
large and there is ``weak'' clustering,
the degree--dependent local clustering coefficient $C(k)$ of Eq.~(\ref{ck}) is given by
\begin{equation}
C(k) = \frac{1}{N \, {\overline k} } \sum_{q, q^\prime} P(q^\prime | k) P(q | k) 
\frac{ P(q^\prime | q) }{P_e(q^\prime) } (q^\prime -1) (q-1)
\end{equation}
which, in terms of the expansion of $P(k^\prime | k)$ in terms of the function
$Q(k^\prime, k)$ through Eq.~(\ref{ansatz}), can be written as
\begin{equation}
C(k) = \frac{1}{N \, {\overline k} } \sum_{q, q^\prime}
P_e(q^\prime) \left[ 1+Q(q^\prime, k) \right]
P_e(q) \left[ 1+Q(q, k) \right]
\left[1+Q(q^\prime, q) \right] (q^\prime -1) (q-1)
\label{ckexp} \end{equation}
The mean clustering coefficient
${\overline C}$ and the clustering coefficient $C$ then follow by Eqs.(\ref{cm}, \ref{c}).
Note that, In the limit of an uncorrelated network ($Q(k^\prime, k)=0$), we have
\begin{equation}
C_{nc} (k) = {\overline C_{nc} } = C_{nc}  = \frac{ (\avg{k}-1)^2}{N \, {\overline k} }
\label{cnk} \end{equation}
Thus, a $k$--dependence of $C(k)$, leading to differences between
$C(k)$, ${\overline C}$, and $C$, signals the presence of nearest--neighbour degree correlations.
\par
One can now use the expansion of $Q(m, n)$ of Eq.~(\ref{qu})
in order to express these clustering coefficients in terms of the expansion parameters
$\beta_{ab}$. We first do so for the local coefficient $C(k)$. 
When expanding Eq.~(\ref{ckexp}), we see that there will be linear, quadratic, and cubic terms
in $Q(m, n)$ present. The linear terms can be evaluated using only the properties
$\avg{u_a} = 0$ and $\avg{u_au_b} = \delta_{ab}$. However, for the quadratic and cubic
terms, we will need to evaluate $\avg{u_a u_b u_1}$. This can be done as follows. 
Let us express the relations of Eqs.~(\ref{gm}, \ref{gma}) relating the
basis vectors $u_a(k)$ and $v_a(k)$ as
\begin{eqnarray}
u_m(k) &=& \sum_{j=1}^m c_{mj} v_j(k) \nonumber\\
v_n(k) &=& \sum_{j=1}^n d_{nj} u_j(k)
\label{uv} \end{eqnarray}
We then have
\begin{equation}
u_a(k) u_1(k) = \sum_{j=1}^a c_{aj} v_j(k) u_1(k) = \sum_{j=1}^a c_{aj} v_j(k) v_1(k)
\label{uprod}\end{equation}
where we have used $u_1(k) = v_1(k)$. Using the definition of $v_n(k)$ in Eq.~(\ref{v}),
we have
\begin{eqnarray}
v_j v_1 &=& \frac{1}{\sqrt{ \sigma_{jj} \sigma_{11} }}
\left[ k^j - \avg{k^j} \right] \left[ k - \avg{k} \right] \nonumber\\
&=&  \frac{1}{\sqrt{ \sigma_{jj} \sigma_{11} }  } \left[
\sqrt{ \sigma_{j+1, j+1} }v_{j+1} -\sqrt{\sigma_{jj}} \avg{k} v_j
-\sqrt{\sigma_{11}} \avg{k^j} v_1 +  \avg{k^{j+1} } -\avg{k}\avg{k^{j} } \right]
\end{eqnarray}
In this we can then use the second relation of Eq.~(\ref{uv}) in this equation
to express the terms involving $v_i$ in terms of $u_i$, and then 
insert this into Eq.(\ref{uprod}) in order to find $u_au_1$ in terms of a series
of terms linear in $u_i$. We find
\begin{eqnarray}
u_a(k)u_1(k) &=& \sum_{j=1}^a \frac{ c_{aj} } { \sqrt{\sigma_{jj} \sigma_{11} }  }\left[
\sqrt{\sigma_{j+1, j+1}  } \sum_{m=1}^{j+1} d_{j+1, m}  u_m(k) \right. \nonumber\\
&-& \left. \sqrt{\sigma_{jj} } \avg{k} \sum_{m=1}^j d_{jm} u_m(k)  - 
\sqrt{\sigma_{11} } \avg{k^j} u_1(k) + \avg{k^{j+1}} - \avg{k^j}\avg{k} \right] 
\end{eqnarray}
We then have
\begin{eqnarray}
\avg{u_au_bu_1} &=&  
\sum_{j=1}^a \frac{ c_{aj} } { \sqrt{\sigma_{jj} \sigma_{11} }  }\left[
\sqrt{\sigma_{j+1, j+1}  } d_{j+1, b} 
-\sqrt{\sigma_{jj} } \avg{k} d_{jb} 
- \sqrt{\sigma_{11} } \avg{k^j} \delta_{b1} \right] \nonumber\\
&\equiv& \gamma_{ab}
\end{eqnarray}
Introducing the notation
\begin{equation}
\avg{ f(q, k) }_q \equiv \sum_q P_e(q) f(q, k)
\end{equation}
for a function $f(q, k)$, 
one can then derive the following relations:
\begin{eqnarray}
\avg{ Q(q, k) u_1(q) }_q &=& k_{nn}^{ (1) }(k) \nonumber\\
\avg{ Q(q, k) k_{nn}^{ (a) }(q) }_q &=& \sum_b \beta_{ab} k_{nn}^{ (b) }(k)  \nonumber\\
\avg{Q(q^\prime, q) Q(q^\prime, k) }_{q^\prime} &=& \sum_a k_{nn}^{ (a) }(q) k_{nn}^{ (a) }(k) 
\nonumber\\
\avg{ Q(q^\prime, q) Q(q^\prime, k) u_1(q^\prime)  }_{q^\prime}  &=&
 \sum_{a, b} \gamma_{ab} k_{nn}^{ (a) }(q) k_{nn}^{ (b) }(k) \nonumber\\
 \avg{ Q(q, k) k_{nn}^{ (a) }(q) u_1(q) }_q &=& 
  \sum_{b, c} \beta_{ac} \gamma_{bc} k_{nn}^{ (b) }(k)
  \end{eqnarray}
Using $u_1(k) = \sqrt{\sigma_{11} }u_1(k) + (\avg{k}-1)$, 
we then find the following contributions to $C(k)$:
\begin{eqnarray}
& &\sum_{q, q^\prime} P_e(q)P_e(q^\prime) (q-1)(q^\prime-1)  = \left[ \avg{k}-1\right]^2 \nonumber\\
 &&\sum_{q, q^\prime} P_e(q)P_e(q^\prime) \left[ Q(q, k) + Q(q^\prime, k) + Q(q, q^\prime) \right]
(q-1)(q^\prime-1) = \sigma_{11}\beta_{11} + 2\sqrt{\sigma_{11}} (\avg{k}-1) k_{nn}^{ (1) }(k)
\nonumber\\
 & &\sum_{q, q^\prime} P_e(q)P_e(q^\prime) \left[Q(q, k)Q(q^\prime, k) 
+Q(q^\prime, k)Q(q, q^\prime) +Q(q, k)Q(q, q^\prime) \right] (q-1)(q^\prime-1) =\nonumber\\
&=& \sigma_{11} \left[ k_{nn}^{ (1) }(k) \right]^2 
+ 2\sqrt{\sigma_{11}} (\avg{k}-1) \sum_a \beta_{1a} k_{nn}^{ (a) }(k) +
2\sigma_{11} \sum_{a, b} \beta_{1b} \gamma_{ab} k_{nn}^{ (a)}(k)\nonumber\\
 & &\sum_{q, q^\prime} P_e(q)P_e(q^\prime) Q(q, q^\prime) Q(q, k)Q(q^\prime, k)
 (q-1)(q^\prime-1) = \sigma_{11} \sum_{a, b, c, d} \beta_{cd} \gamma_{ac}\gamma_{bd}
  k_{nn}^{ (a)}(k)  k_{nn}^{ (b)}(k) \nonumber\\
  &+& 2\sqrt{\sigma_{11}} (\avg{k}-1) \sum_{a, b, c} \beta_{ac}\gamma_{bc} 
    k_{nn}^{ (a)}(k)  k_{nn}^{ (b)}(k)  +
 ( \avg{k}-1)^2 \sum_{ab} \beta_{ab}  k_{nn}^{ (a)}(k)  k_{nn}^{ (b)}(k) 
\end{eqnarray}
We then find the local clustering coefficient $C(k)$, normalized to the non--correlated
value $C_{nc} $ of Eq.~(\ref{cnk}), can be written as
\begin{equation}
\frac{ C(k) } { C_{nc} } = 1 + F + \sum_a G_a k_{nn}^{ (a)}(k)  +
\sum_{a, b} H_{a b} k_{nn}^{ (a)}(k) k_{nn}^{ (b)}(k)
\label{ck1} \end{equation}
where
\begin{eqnarray}
F &=& \frac{ \sigma_{11} \beta_{11}  } { (\avg{k}-1)^2 } \nonumber\\
G_a &=& \frac{2 \sqrt{\sigma_{11} } } { (\avg{k}-1)  }
\left[ \delta_{1a} + \beta_{1a} \right] +  \frac{2 \sigma_{11}  } { (\avg{k}-1)^2  }
 \sum_b \beta_{1b} \gamma_{ab} \nonumber\\
 H_{ab} &=& \beta_{ab} +  \frac{ 2 \sqrt{\sigma_{11} } } { (\avg{k}-1) }
 \sum_c \beta_{ac} \gamma_{bc} +
   \frac{ \sigma_{11}  } { (\avg{k}-1)^2  } \left[ \delta_{1a} \delta_{1b} +
  \sum_{c, d} \beta_{cd} \gamma_{ac}\gamma_{bd} \right]
   \end{eqnarray}
To calculate the clustering coefficient $C$ of Eq.~(\ref{c}), we will need
\begin{eqnarray}
 \avg{ k_{nn}^{ (a) }(k) u_1(k) }_k &=& \beta_{a1} \nonumber\\
 \avg{ k_{nn}^{ (a) }(k) k_{nn}^{ (b) }(k) }_k &=& \sum_c \beta_{ac} \beta_{bc} \nonumber\\
  \avg{ k_{nn}^{ (a) }(k) k_{nn}^{ (b) }(k) u_1(k) }_k &=& \sum_{c, d} \beta_{ac}\beta_{bd} \gamma_{cd}
\end{eqnarray}
This leads to
\begin{equation}
\frac{C}{C_{nc} } = 1 + F + \frac{ \sqrt{\sigma_{11} } } { (\avg{k} - 1) } \sum_a G_a \beta_{a1}
+ \sum_{a, b, c} H_{ab} \beta_{ac} \beta_{bc} + 
 \frac{ \sqrt{\sigma_{11} } } { (\avg{k} - 1) } \sum_{a, b, c, d} H_{ab} \beta_{ac} \beta_{bd}\gamma_{cd}
\label{c1}\end{equation}
\par
In order to get a qualitative sense of the behaviour of $C(k)$ of Eq.~(\ref{ck1}) and
$C$ of Eq.~(\ref{c1}), let us assume that the expansion parameters $\beta_{ab}$ are
small, such that we need only keep terms up to linear order in them. Since by the definition of
Eq.~(\ref{annd}) the generalized ANND
$k_{nn}^{ (a) }(k)$ is linear in $\beta_{ab}$, we find the local clustering coefficient
$C(k)$ of Eq.~(\ref{ck1}) could be approximated as
\begin{equation}
\frac{ C(k)} {C_{nc} } \approx 1 + \frac{ \sigma_{11} \beta_{11} } { (\avg{k} - 1)^2 }
+ \frac{ 2 \sqrt{\sigma_{11} }} { (\avg{k} - 1) }  k_{nn}^{ (1) }(k)
 \label{ckapp}\end{equation}
while the corresponding approximation of the clustering coefficient $C$ of Eq.~(\ref{c1}) is
\begin{equation}
\frac{C}{C_{nc}} \approx 1 + \frac{ 3 \sigma_{11} \beta_{11}  } {  (\avg{k} - 1)^2 }
\label{capp}\end{equation}
In the next section we consider a couple of examples to illustrate the
behaviour of various quantities discussed in this and the previous sections.
The full numerical results for the clustering coefficients suggest that, for 
for small $\beta_{ab}$, this linear approximation just considered
gives at least give a fairly good qualitative picture
for how the clustering coefficients behave. In particular, from Eq.~(\ref{ckexp}), the local
degree--dependent clustering coefficient $C(k)$ has a $k$--dependence that
follows approximately
that of the first--order ANND $k_{nn}^{(1)}(k)$ (compare Fig.~\ref{knn-pos} and Fig.~\ref{ck-pos},
and also Fig.~\ref{knn-neg} and Fig.~\ref{ck-neg}). As well, the clustering coefficient $C$ of
Eq.~(\ref{capp}) is increased/decreased slightly, relative to the non--correlated value,
for  positive/negative $\beta_{11}$, corresponding to assortative/disassortative mixing (see
the discussion immediately following, respectively, Fig.~\ref{ck-pos} and Fig.~\ref{ck-neg}).
\section{Examples}
\label{examples}
\begin{figure}[H]
\begin{center}
\includegraphics[width=4.5in]{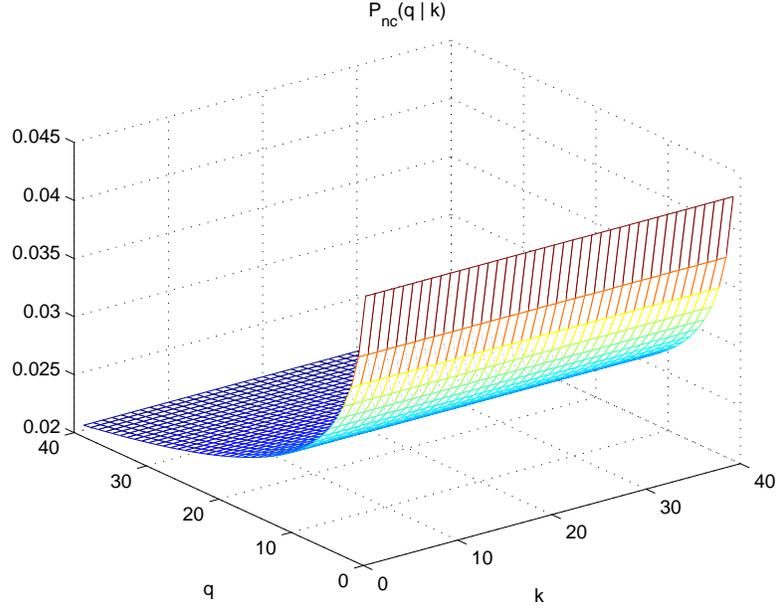}
\caption{The uncorrelated conditional probability $P_{nc}(q | k)$ for $P(k) \sim k^\gamma$, with 
$\gamma = -1.2$.}
\label{pnc}
\end{center}
\end{figure}

In order to illustrate the relative size of the effects considered in the previous sections, in this
Section we examine two examples of explicit choices of the parameters
$\beta_{ab}$ in the expansion of $Q(k^\prime, k)$ of Eq.~(\ref{ansatz}). In keeping with the spirit of a perturbative approach, we expect that, if we keep relatively few terms in the expansion,
 the corrections to the uncorrelated case 
$Q(k^\prime, k) = 0$ will be small in some sense.
In both cases we use a degree distribution function $P(k) \sim k^\gamma$, with 
$\gamma = -1.2$, and consider degrees $1 \le k \le 40$. The uncorrelated joint probability $P_{nc}(q | k) = P_e(q)$
appears as in Fig.~\ref{pnc}.
\par
\begin{figure}[H]
\begin{center}
\includegraphics[width=4.5in]{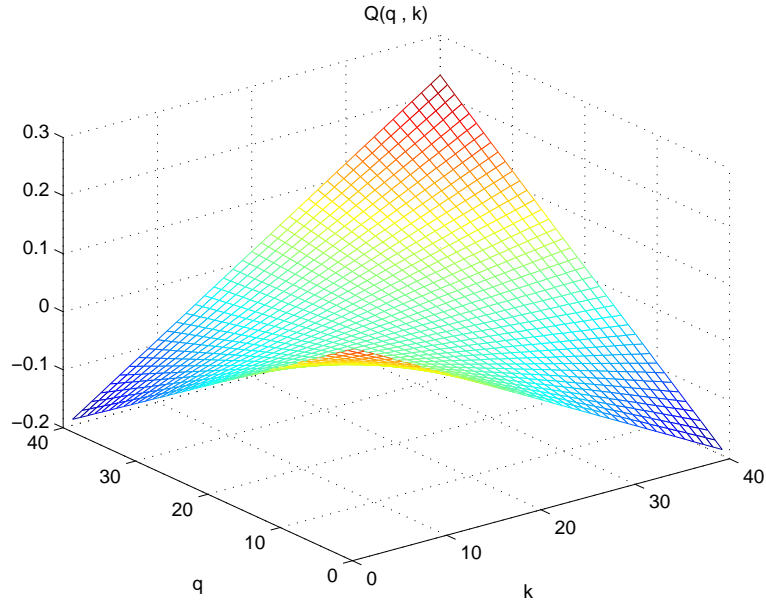}
\caption{The expansion function $Q(q, k)$ for $\beta_{11} = 0.07$.}
\label{q-pos}
\end{center}
\end{figure}
\begin{figure}[H]
\begin{center}
\includegraphics[width=4.5in]{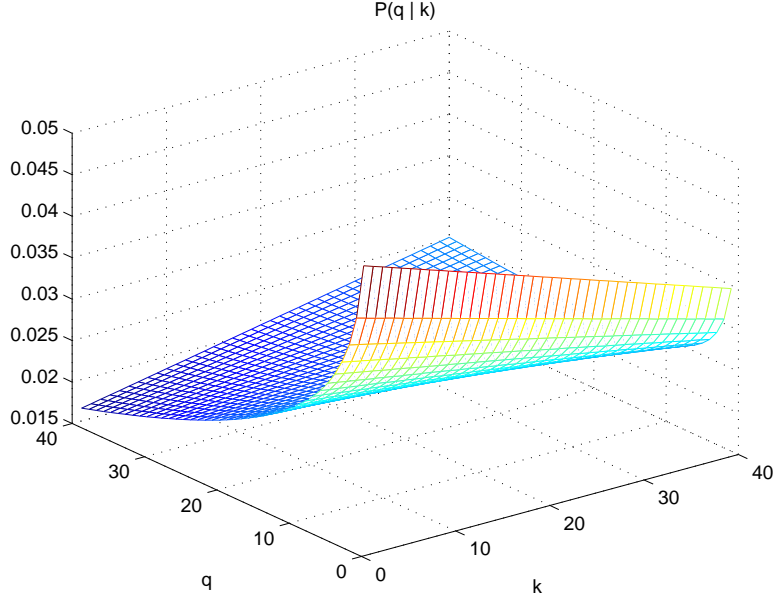}
\caption{The conditional probability $P(q | k)$ for $\beta_{11} = 0.07$.}
\label{pcon-pos}
\end{center}
\end{figure}
\begin{figure}[H]
\begin{center}
\includegraphics[width=4.5in]{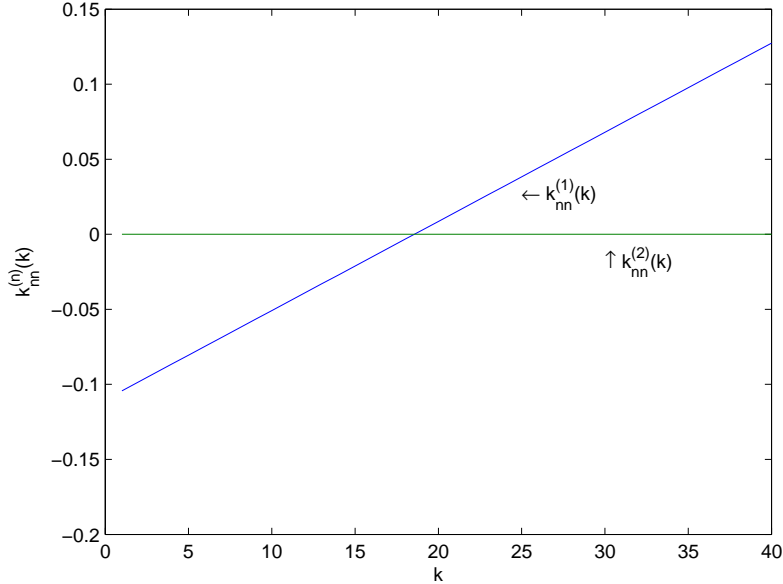}
\caption{The first--order ANND $k_{nn}^{(1)}(k)$ for $\beta_{11} = 0.07$.}
\label{knn-pos}
\end{center}
\end{figure}
\begin{figure}[H]
\begin{center}
\includegraphics[width=4.5in]{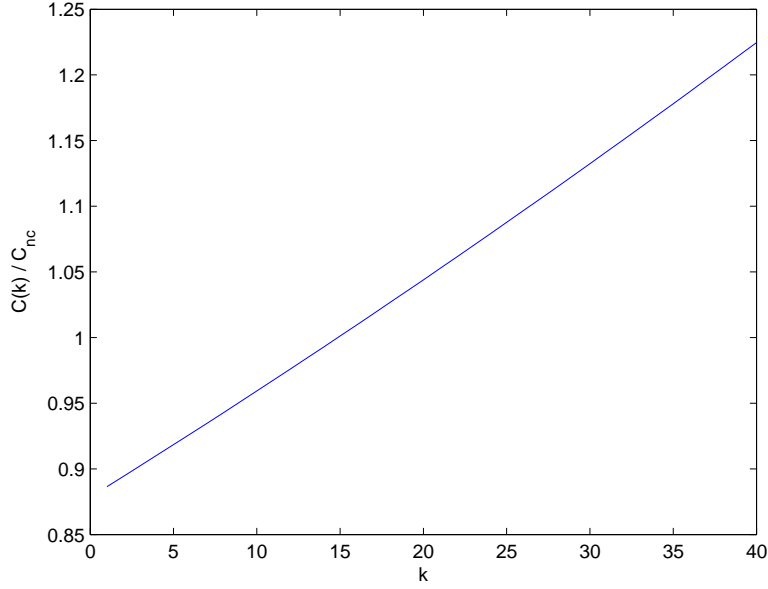}
\caption{The degree--dependent local clustering coefficient $C(k) / C_{nc}$ for $\beta_{11} = 0.07$,
normalized to the uncorrelated value.}
\label{ck-pos}
\end{center}
\end{figure}

The first case will be one exhibiting assortative mixing. For this we choose a single
linear term $\beta_{11} = 0.07$. The function $Q(q, k)$ appears in Fig.~\ref{q-pos},
while the conditional probability $P(q | k) = P_e(q)[ 1 + Q(q, k)]$ appears in
Fig.~\ref{pcon-pos}.
As expected with assortative mixing, we see from the conditional probability of 
Fig.~\ref{pcon-pos} that there is an increased likelihood of nodes with the same
degree to be connected. Fig.~\ref{knn-pos} shows the generalized ANND of
Eq,(\ref{annd}); in this case, only $k_{nn}^{(1)}(k)$ will be non--zero, and it is a linear 
function.
The only non--zero correlation coefficient $r_{ab}$ of Eq.~(\ref{correlation}) is $r_{11}=1$,
as expected, since the linear correlation is perfect in this case. Finally, in Fig.~\ref{ck-pos}
we plot the degree--dependent local clustering coefficient $C(k)$ of Eq.~(\ref{ck}), which
has been normalized to the uncorrelated value $C_{nc}$ of Eq.~(\ref{cnk}).
Corresponding to this $C(k)$, we find the mean clustering coefficient ${\overline C}$
of Eq.~(\ref{cm}), normalized to $C_{nc}$, to be ${\overline C} / C_{nc} = 0.94$, while the  
clustering coefficient $C$ of Eq.~(\ref{c}), also normalized to $C_{nc}$, to be
$C / C_{nc} = 1.10$; this represents about a 10\% change from the uncorrelated state.
\par
\begin{figure}[H]
\begin{center}
\includegraphics[width=4.5in]{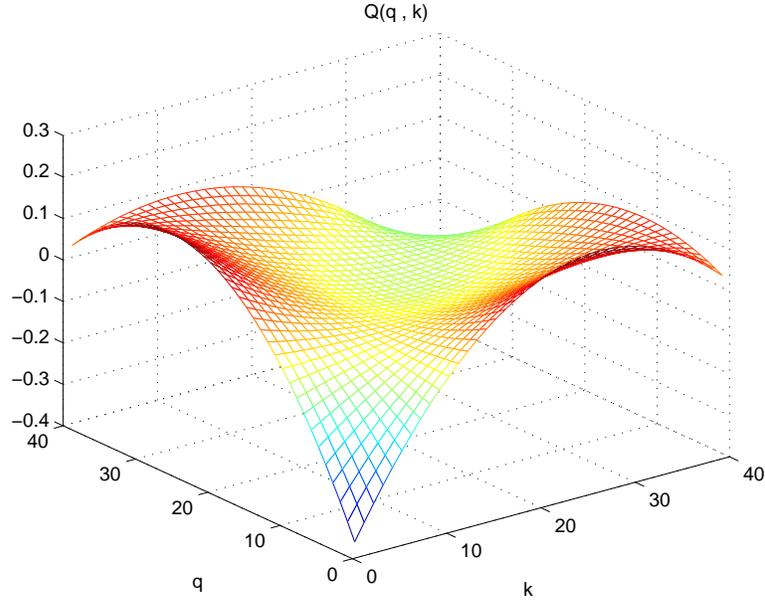}
\caption{The expansion function $Q(q, k)$ for $\beta_{11} = -0.07$, 
$\beta_{12} = 0.02 = \beta_{21}$, and $\beta_{22} = -0.04$.}
\label{q-neg}
\end{center}
\end{figure}
\begin{figure}[H]
\begin{center}
\includegraphics[width=4.5in]{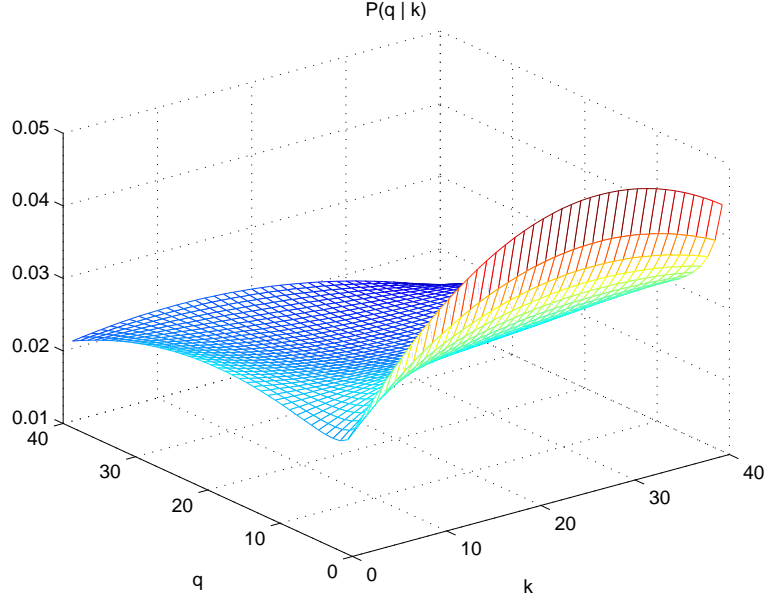}
\caption{The conditional probability $P(q | k)$ for $\beta_{11} = -0.07$, 
$\beta_{12} = 0.02 = \beta_{21}$, and $\beta_{22} = -0.04$.}
\label{pcon-neg}
\end{center}
\end{figure}
\begin{figure}[H]
\begin{center}
\includegraphics[width=4.5in]{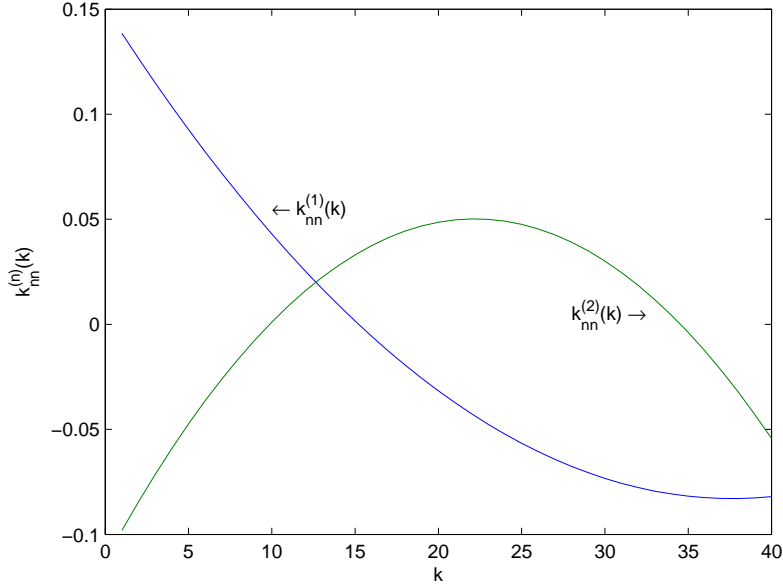}
\caption{The first--order ANND $k_{nn}^{(1)}(k)$ and second--order ANND
 $k_{nn}^{(2)}(k)$ for $\beta_{11} = -0.07$, 
$\beta_{12} = 0.02 = \beta_{21}$, and $\beta_{22} = -0.04$.}
\label{knn-neg}
\end{center}
\end{figure}
\begin{figure}[H]
\begin{center}
\includegraphics[width=4.5in]{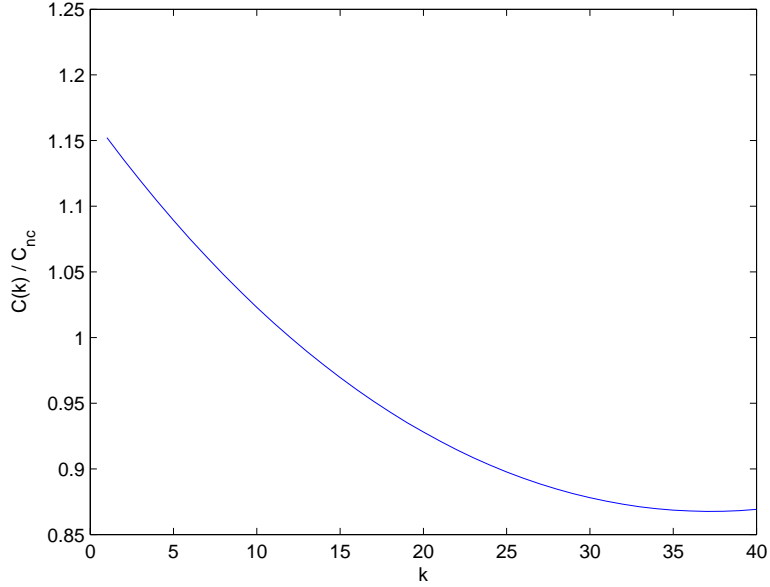}
\caption{The degree--dependent local clustering coefficient $C(k) / C_{nc}$ for $\beta_{11} = -0.07$, 
$\beta_{12} = 0.02 = \beta_{21}$, and $\beta_{22} = -0.04$, 
normalized to the uncorrelated value.}
\label{ck-neg}
\end{center}
\end{figure}

The second case we examine will be one exhibiting disassortative mixing. For this we consider
including both linear and quadratic terms in the expansion of $Q(q, k)$, and we
choose $\beta_{11} = -0.07$, $\beta_{12} = 0.02 = \beta_{21}$, and $\beta_{22} = -0.04$.
The function $Q(q, k)$ appears in Fig.~\ref{q-neg},
while the conditional probability $P(q | k) = P_e(q)[ 1 + Q(q, k)]$ appears in
Fig.~\ref{pcon-neg}.
As expected with disassortative mixing, in this case we see from the conditional probability of 
Fig.~\ref{pcon-neg} that there is an increased likelihood of nodes of different
degree to be connected. Fig.~\ref{knn-neg} shows the generalized ANND of
Eq,(\ref{annd}); in this case, both  $k_{nn}^{(1)}(k)$ and $k_{nn}^{(2)}(k)$
will be non--zero.
It is found that the only non--zero correlation coefficients $r_{ab}$ of Eq.~(\ref{correlation}) are
 $r_{11}=-0.96$, $r_{12}=0.27$, $r_{21}=0.45$, and $r_{22}=-0.89$.
Finally, in Fig.~\ref{ck-neg}
we plot the degree--dependent local clustering coefficient $C(k)$ of Eq.~(\ref{ck}), which
has been normalized to the uncorrelated value $C_{nc}$ of Eq.~(\ref{cnk}).
Corresponding to this $C(k)$, we find the mean clustering coefficient ${\overline C}$
of Eq.~(\ref{cm}), normalized to $C_{nc}$, to be ${\overline C} / C_{nc} = 1.08$, while the  
clustering coefficient $C$ of Eq.~(\ref{c}), also normalized to $C_{nc}$, to be
$C / C_{nc} = 0.91$; this represents about a 10\% change from the uncorrelated state.
\section{Conclusions}
\label{conclusions}
We have examined an expansion of the conditional probability 
$P(k^\prime, k) = P_e(k^\prime)[ 1 + Q(k^\prime, k)]$ about
the uncorrelated case $P_{nc}(k^\prime | k) =  P_e(k^\prime)$ in terms of 
symmetric polynomials in $k^\prime$ and $k$. Setting aside the question of
convergence, we find a systematic expansion is possible, and will involve
expansion coefficients $\beta_{ab}$. Having specified these coefficients up to
a certain order, the usual measures of nearest--neighbour degree correlations  --
the Average Nearest Neighbour Degree ANND, Pearson--inspired correlation coefficients, 
and various clustering coefficients -- can be calculated. In the present case, since
non--linear terms in the expansion would in principle appear, appropriate
generalizations of these measures of degree correlations were introduced.
\par
One possible use for the type of expansion discussed in this paper might be
as a means to estimate qualitatively the effect of nearest--neighbour degree correlations in models 
describing the evolution of states of nodes on specific networks -- for example, the propagation of disease on the network. This can be formulated in terms of the equations governing the evolution of probabilities.. This can come by formulating
the equations governing the evolution in terms of probabilities. To
see this in a general sense, let $\rho_{X, k}$ be the probability that a node of degree $k$ is
in a state $X$ at time $t$. A differential equation describing the time evolution of
$\rho_{X, k}$ might then contain terms such as
\begin{equation}
\frac {d\, \rho_{X, k} } {dt} = g \rho_{X, k} \rho_{Y, k} + \ldots
\end{equation}
which would describe how the transition $X \to Y$ in the system, parameterized by a rate $g$,
affects the probability $\rho_{X, k}$. Nearest--neighbour degree correlations could
be incorporated into this model by consideration of the interaction term  \cite{complex}
\begin{equation}
\frac {d\, \rho_{X, k} } {dt} = g k \rho_{X, k} \Theta_{Y, k} + \ldots
\end{equation}
where
\begin{equation}
\Theta_{Y, k} = \sum_{k^\prime} \frac{k^\prime -1}{k^\prime} P(k^\prime | k) \rho_{Y, k^\prime}
\end{equation}
 is the probability that a neighbour of the node, chosen randomly from amongst its
$k$ neighbours, is in a state $Y$. Thus, specifying a conditional probability function $P(k^\prime | k)$
that differs from the uncorrelated case $P_{nc}(k^\prime | k) = P_e(k^\prime)$ would allow
one to see the effects of different types of degree correlations in this model.
\par
If one is to use the expansion of $Q(k^\prime, k)$ of Eq.~(\ref{qu}) in specifying a conditional
probability $P(k^\prime | k)$ containing correlations, one must decide on
the values of the expansion parameters
$\beta_{ab}$ to use, and at what point does one know that enough terms have been
kept. A mild constraint on the expansion is that, being a perturbative expansion,
the corrections about the unperturbed case should be small, and that inclusion of higher--order
effects should not affect significantly the results of the presumably more important
lower--order terms. An equivalent statement of this is that, for a given set of
parameters, changing them slightly would not change the overall qualitative picture.
It may be possible, by examining classes of real networks,
to be able to say something about the relative magnitude of the various $\beta_{ab}$
coefficients. However, if one wanted just a qualitative estimate of the relative effects
of such degree correlations in the model under consideration, then one could consider
a ``small'' number of terms in this expansion with ``small'' values of the $\beta_{ab}$
parameters, where ``small'' in this context is defined through the constraint that
the effects do not significantly alter the uncorrelated case. The examples of
Section \ref{examples} show that a reasonable set of parameters can be chosen
which incorporate various types of degree correlations. Such an approach 
would not allow one to say anything quantitative about a real network, but it would allow
one to decide, with some degree of confidence, whether or not inclusion of degree
correlations in the model might lead to significant effects, and thus would be worthy
of further, more detailed, study.
\acknowledgments
This work was supported by the Natural Sciences and Engineering Research
Council of Canada.

\end{document}